\documentstyle[aps,prl,epsfig,preprint,citeup]{revtex}
\renewcommand{\vec}[1]{\relax\ifmmode\mathchoice
{\mbox{\boldmath$\relax\displaystyle#1$}}
{\mbox{\boldmath$\relax\textstyle#1$}}
{\mbox{\boldmath$\relax\scriptstyle#1$}}
{\mbox{\boldmath$\relax\scriptscriptstyle#1$}}\else
\hbox{\boldmath$\relax\textstyle#1$}\fi}
\begin{document}
\pagestyle{myheadings}
\markboth{Helbing/Farkas/Vicsek: Simulating Dynamical Features
of Escape Panic}
{Helbing/Farkas/Vicsek: Simulating Dynamical  Features
of Escape Panic}
\title{\mbox{}\\[-2.2cm]Simulating Dynamical Features
of Escape Panic\\[-0cm]}
\author{Dirk Helbing$^{*,+}$,
Ill\'{e}s Farkas,$^\ddag$
and Tam\'{a}s Vicsek$^{*,\ddag}$}
\address{$^*$ Collegium Budapest~-- Institute for Advanced Study,
Szenth\'{a}roms\'{a}g u. 2,\\
H-1014 Budapest, Hungary\\
$^+$ Institute for Economics and Traffic, Dresden University of
Technology,\\
D-01062 Dresden, Germany\\
$^\ddag$ Department of Biological Physics, E\"otv\"os University,\\
P\'azm\'any P\'eter S\'et\'any 1A, H-1117 Budapest, Hungary\\
{\tt helbing@trafficforum.de; fij@elte.hu;
vicsek@angel.elte.hu}\\[8mm]}
\maketitle
\draft
\par
{\bf One of the most disastrous forms of collective human behaviour is the kind
of crowd stampede induced by panic, often leading to fatalities as
people are crushed or trampled.
Sometimes this behaviour is triggered in
life-threatening situations such as fires in crowded 
buildings;\cite{Fire1,Fire2} at other
times, stampedes can arise from the rush for seats\cite{Hills,Stadion} 
or seemingly without causes. 
Tragic examples within
recent months include the panics in Harare, Zimbabwe, and
at the Roskilde rock concert in Denmark.
Although engineers are finding ways to alleviate the
scale of such disasters, their frequency 
seems to be increasing with the number and size of mass events.\cite{Fire2,calibrate} Yet,
systematic studies of panic behaviour,\cite{Exp}$^-$\cite{Coll2}
and quantitative theories capable of
predicting such crowd
dynamics,\cite{calibrate,FireSim2}$^-$\cite{FireSim4} 
are rare. Here we show that simulations
based on a model of pedestrian behaviour can provide valuable insights into
the mechanisms of and preconditions for panic and 
jamming by incoordination. Our results
suggest practical ways of minimising the harmful consequences of such
events and the existence of an optimal escape strategy,
corresponding to a suitable mixture of individualistic
and collective behaviour.}
\par
Up to now, panics as a particular form of collective behaviour occuring
in situations of scarce or dwindling resources\cite{Fire1,Exp} has been
mainly studied from the perspective of 
social psychology.\cite{Exp1}$^-$\cite{Coll2}
Panicking individuals tend to show maladaptive and relentless mass behaviour
like jamming and life-threatening 
overcrowding,\cite{Fire1}$^-$\cite{Stadion,SocPsy1}
which has often been attributed to social
contagion\cite{Fire1,Stadion,SocPsy1}
(see Brown\cite{Coll2} for
an overview of theories). According to Mintz,\cite{Exp} 
the observed jamming is a result of
incoordination and depends on the reward structure.
\par
After having carefully studied the related 
socio-psychological literature,\cite{Exp}$^-$\cite{Coll2}
reports in the media and available video materials 
(see {\tt http://angel.elte. hu/$\widetilde{\hphantom{n}}$panic/}),
empirical  investigations,\cite{Fire1,Fire2,Hills}
and engineering handbooks,\cite{eng1,eng2} we can summarise the following
characteristic features of escape panics:
(i) People move or try to move considerably faster than normal.\cite{eng1}
(ii) Individuals start pushing, and interactions among people 
become physical in nature.
(iii) Moving and, in particular, passing of a bottleneck 
becomes incoordinated.\cite{Exp}
(iv) At exits, arching and clogging are observed.\cite{eng1}
(v) Jams are building up.\cite{Exp1}
(vi) The physical interactions in the jammed crowd 
add up and cause dangerous pressures up to
4,450 Newtons per meter,\cite{Fire2,calibrate} which can bend steel barriers or
tear down brick walls.
(vii) Escape is further slowed down by fallen or injured people turning
into ``obstacles''.
(viii) People show a tendency of mass behaviour, i.e., 
to do what other people do.\cite{Fire1,SocPsy1}
(ix) Alternative exits are often overlooked or not efficiently used
in escape situations.\cite{Fire1,Fire2}
\par
These observations have encouraged us to model
the collective phenomenon of escape panic in the spirit of self-driven
many-particle systems.
Our computer simulations of the crowd dynamics of pedestrians
are based on a generalised force model,\cite{Freezing} which is
particularly suited to describe the fatal build up of pressure observed during
panics.\cite{Fire2}$^-$\cite{Stadion,calibrate} We assume a mixture of
socio-psychological\cite{Pedestrians} and physical forces influencing
the behaviour in a crowd: Each of $N$ pedestrians $i$ of mass $m_i$ likes
to move with a certain desired speed $v_i^0$ into a certain direction
$\vec{e}_i^0$, and therefore tends to correspondingly
adapt his or her actual velocity $\vec{v}_i$ with a certain
characteristic time $\tau_i$. Simultaneously, he or she tries to keep
a velocity-dependent distance to
other pedestrians $j$ and walls $W$. 
This can be modelled by ``interaction forces'' $\vec{f}_{ij}$ and
$\vec{f}_{iW}$, respectively.
In mathematical terms, the change of velocity in time $t$ is then
given by the acceleration equation
\begin{equation}
 m_i \frac{d\vec{v}_i}{dt} =
 m_i \frac{v_i^0(t) \vec{e}_i^0(t) - \vec{v}_i(t)}{\tau_i}
 + \sum_{j (\ne i)} \vec{f}_{ij} + \sum_W \vec{f}_{iW}    \, ,
\end{equation}
while the change of position $\vec{r}_i(t)$
is given by the velocity $\vec{v}_i(t) = d\vec{r}_i/dt$.
We describe the {\em psychological tendency} of two pedestrians $i$ and $j$
to stay away from each other by a repulsive interaction force
$A_i\exp[(r_{ij}-d_{ij})/B_i] \, \vec{n}_{ij}$,
where $A_i$ and $B_i$ are constants.
$d_{ij} = \|\vec{r}_i - \vec{r}_j\|$ denotes the distance between the
pedestrians' centers of mass, and
$\vec{n}_{ij}= (n_{ij}^1,n_{ij}^2)
= (\vec{r}_i-\vec{r}_j)/d_{ij}$ is the normalised vector
pointing from pedestrian $j$ to $i$. If their distance $d_{ij}$ is
smaller than the sum $r_{ij} = (r_i +
r_j)$ of their radii $r_i$ and $r_j$, the pedestrians touch each other.
In this case, we assume two additional forces
inspired by granular interactions,\cite{Herrmann,Granular} which are essential
for understanding the particular effects in panicking crowds: a
{\em ``body force''} $k(r_{ij}-d_{ij})\, \vec{n}_{ij}$ counteracting body compression
and a {\em ``sliding friction force''} $\kappa (r_{ij} - d_{ij})\, \Delta v_{ji}^t \,
\vec{t}_{ij}$ impeding {\em relative} tangential motion, if pedestrian $i$ comes close to
$j$. Herein, $\vec{t}_{ij} = (-n_{ij}^2, n_{ij}^1)$ means the tangential
direction and $\Delta v_{ji}^t = (\vec{v}_j -\vec{v}_i) \cdot \vec{t}_{ij}$
the tangential velocity difference, while $k$ and $\kappa$ represent large
constants. In summary, we have
\begin{equation}
 \vec{f}_{ij} = \left\{ A_i\exp[(r_{ij}-d_{ij})/B_i]
 + k g(r_{ij}-d_{ij}) \right\} \vec{n}_{ij}
 + \kappa g(r_{ij}-d_{ij}) \Delta v_{ji}^t \, \vec{t}_{ij} \, ,
\end{equation}
where the function $g(x)$ is zero, if the pedestrians do not touch each
other ($d_{ij} > r_{ij}$), otherwise equal to the argument $x$.
\par
The interaction with the
walls is treated analogously, i.e., if
$d_{iW}$ means the distance to wall $W$,
$\vec{n}_{iW}$ denotes the direction perpendicular to it, and
$\vec{t}_{iW}$ the direction tangential to it,
the corresponding interaction force with the wall reads
\begin{equation}
 \vec{f}_{iW} = \left\{ A_i \exp[(r_{i}-d_{iW})/B_i]
 + k g(r_i-d_{iW}) \right\} \vec{n}_{iW}
 - \kappa g(r_i-d_{iW}) (\vec{v}_i\cdot\vec{t}_{iW}) \, \vec{t}_{iW} \, .
\end{equation}
\par
Probably due to the fact that escape panics are 
unexpected and dangerous events, which
also excludes real-life experiments, we could not find 
suitable data on escape panics to test our model {\em quantitatively}. 
This scarcity of data calls for reliable models. We have, therefore, 
specified the parameters as follows: With a mass of
$m_i=80$~kg, we represent an average soccer fan. The desired velocity
$v_i^0$ can reach more than 5~m/s (up to 10~m/s),\cite{eng2} but the observed 
free velocities for leaving a room correspond to  
$v_i^0 \approx 0.6$~m/s under relaxed,
$v_i^0 \approx 1$~m/s under normal, and $v_i^0 \lesssim 1.5$~m/s
under nervous conditions.\cite{eng1} A
reasonable estimate for the acceleration time is $\tau_i = 0.5$~s. 
With $A_i = 2\cdot 10^3$~N and $B_i = 0.08$~m
one can reflect the distance kept at normal desired velocities\cite{eng2} 
and fit the measured flows through bottlenecks\cite{eng2}, amounting to
0.73 persons per second for an effectively 1~meter 
wide door under conditions with $v_i^0 \approx 0.8$~m/s.
The parameters $k=1.2\cdot 10^5$~kg\,s$^{-2}$ 
and $\kappa = 2.4\cdot 10^5$~kg\,m$^{-1}$s$^{-1}$ determine the
obstruction effects in cases of physical interactions. 
Although, in reality, most parameters are varying individually, we chose
identical values for all pedestrians to minimise the number
of parameters for reasons of calibration and robustness, and to
exclude irregular outflows because of parameter variations. However,
to avoid model artefacts (gridlocks by exactly balanced forces in
symmetrical configurations), a small amount of irregularity 
of almost arbitrary kind is needed. 
This irregularity was introduced by uniformly distributed
pedestrian diameters $2r_i$ in the interval $[0.5\mbox{ m},0.7\mbox{ m}]$, 
approximating the distribution of shoulder widths of soccer fans.   
\par
Based on the above model assumptions, we will now simulate several
important phenomena of escape panic, which are insensitive to
reasonable parameter variations, but fortunately become less pronounced for wider exits.
\par
{\it 1. Transition to Incoordination due to Clogging}.
The simulated outflow from a room is well-coordinated and regular, if the
desired velocities $v_i^0=v_0$ are normal. However, for desired
velocities above 1.5~m/s, i.e. for people in a rush, 
we find an irregular
succession of arch-like blockings of the exit and avalanche-like
bunches of leaving pedestrians, when the arches break (see
Fig.~\ref{fig2}{\sf a, b}). 
This phenomenon is compatible with the empirical
observations mentioned above and comparable to intermittent clogging
found in granular flows through funnels or hoppers\cite{Herrmann,Granular}
(although this has been attributed to {\em static} friction between particles
without remote interactions, and the transition to clogging has been
observed for small enough openings rather than for a variation of
the driving force).
\par
{\em 2. ``Faster-Is-Slower Effect'' due to Impatience}.
Since clogging is connected with delays,
trying to move faster (i.e., increasing $v_i^0$) can cause a smaller
average speed of leaving, if the friction
parameter $\kappa$ is large enough (see Fig.~\ref{fig2}{\sf c, d}). 
This ``faster-is-slower effect'' is particularly tragic in the
presence of fires, where fleeing people reduce their own chances
of survival. The related fatalities can be estimated by the number
of pedestrians reached by the fire front (see {\tt
http://angel.elte.hu/$\widetilde{\hphantom{n}}$panic/}).
\par
Since our friction term has, on average, no deceleration effect
in the crowd, if the walls are sufficiently remote,
the arching underlying the clogging effect requires a {\em combination} of
several effects:
i) slowing down due to a bottleneck such as a door and
ii) strong inter-personal friction, which becomes dominant when pedestrians get
too close to each other. Consequently, the danger of clogging can
be minimised by avoiding bottlenecks in the construction of
stadia and public buildings. Notice, however, that jamming can also
occur at widenings of escape routes! This surprising result is
illustrated in Fig.~\ref{fig3}. {\em Improved
outflows can be reached by columns placed asymmetrically in front of the exits,
which also prevent the build up of fatal pressures}
(see {\tt http://angel.elte.hu/$\widetilde{\hphantom{n}}$panic/}).
\par
{\it 3. Mass Behaviour.} 
We investigate a situation in which pedestrians are
trying to leave a smoky room, but first have to find one of the
invisible exits (see Fig.~\ref{fig4}{\sf a}). Each pedestrian $i$ may
either select an individual direction $\vec{e}_i$ or 
follow the average direction $\langle \vec{e}_j^0(t) \rangle_i$
of his neighbours $j$ in a certain radius $R_i$,\cite{Birds} or
try a mixture of
both. We assume that both options are weighted with some parameter $p_i$:
\begin{equation}
 \vec{e}_i^0(t)
= {\cal N}\left[ \left(1- p_i \right) \vec{e}_i
+ p_i  \, \langle \vec{e}_j^0(t) \rangle_i \right] \, ,
\end{equation}
where ${\cal N}(\vec{z}) = \vec{z} / \|\vec{z}\|$
denotes normalisation of a vector
$\vec{z}$. As a consequence, we have individualistic behaviour
if $p_i$ is low, but herding behaviour if $p_i$ is high. Therefore,
$p_i$ reflects the degree of panics of individual~$i$.
\par
Our model suggests that neither
individualistic nor herding behaviour performs well
(see Fig.~\ref{fig4}{\sf b}).
Pure individualistic behaviour means that each
pedestrian finds an exit only accidentally, while pure herding
behaviour implies that the complete crowd is eventually moving into
the same and probably blocked
direction, so that available exits are not efficiently used,
in agreement with observations. According to
Figs.~\ref{fig4}{\sf b} and {\sf c}, we expect optimal chances of survival for a certain
mixture of individualistic and herding behaviour, where 
individualism allows some people to detect the exits and herding guarantees
that successful solutions are imitated by the others. 
If pedestrians follow the walls instead of ``reflecting'' at them, 
we expect that herd following causes jamming and inefficient use 
of doors as well (see Fig.~\ref{fig2}), while
individualists moving in opposite directions obstruct each other.
\par
In summary, we have developed a continuous pedestrian model based on
plausible interactions, which is, due to its simplicity, robust with respect
to parameter variations and suitable for drawing conclusions about the 
possible mechanisms beyond escape panic (regarding an increase of
the desired velocity, strong friction effects
during physical interactions, and herding). After having calibrated the 
model parameters to available data on pedestrian flows, we
managed to reproduce many observed phenomena including i) the build up of
pressure, ii) clogging effects at bottlenecks, iii) jamming at widenings,
iv) the ``faster-is-slower effect'', v) inefficient use of alternative
exits (see Fig.~\ref{fig4}{\sf d}), and vi) initiation
of panics by counterflows and impatience (i.e., seemingly without
any logical reason, see Fig.~\ref{fig2}). Moreover, we are able to simulate situations
of dwindling resources and estimate the casualties. Therefore,
the model can be used to test buildings for their suitability in
emergency situations.
It accounts for both, the different dynamics in normal
and panic situations just by changing a single parameter $p_i=p$.
In addition, our simulations suggest that the optimal behaviour in escape
situations is a suitable mixture of individualistic and herding behaviour.
\par
We are now calling for complementary data and additional video material
on escape panics 
to test our model quantitatively and compare it with alternative
ones which, for example, include
direction- and velocity-dependent interpersonal
interactions, specify the individual variation of
parameters, study the effect of fluctuations, 
consider falling people, integrate acoustic information
exchange, implement more complex strategies and interactions (also
three-dimensional ones), or
allow for switching of strategies. A superior theory would have to
reproduce the empirical findings equally well with less parameters,
reach a better quantitative agreement with data with the same number,
or reflect additional observations.

{\em Acknowledgments:}
D.H. thanks the German Research Foundation (DFG) for financial support
by a Heisenberg scholarship. T.V. and I.F. are grateful for
partial support by OTKA and FKFP. 
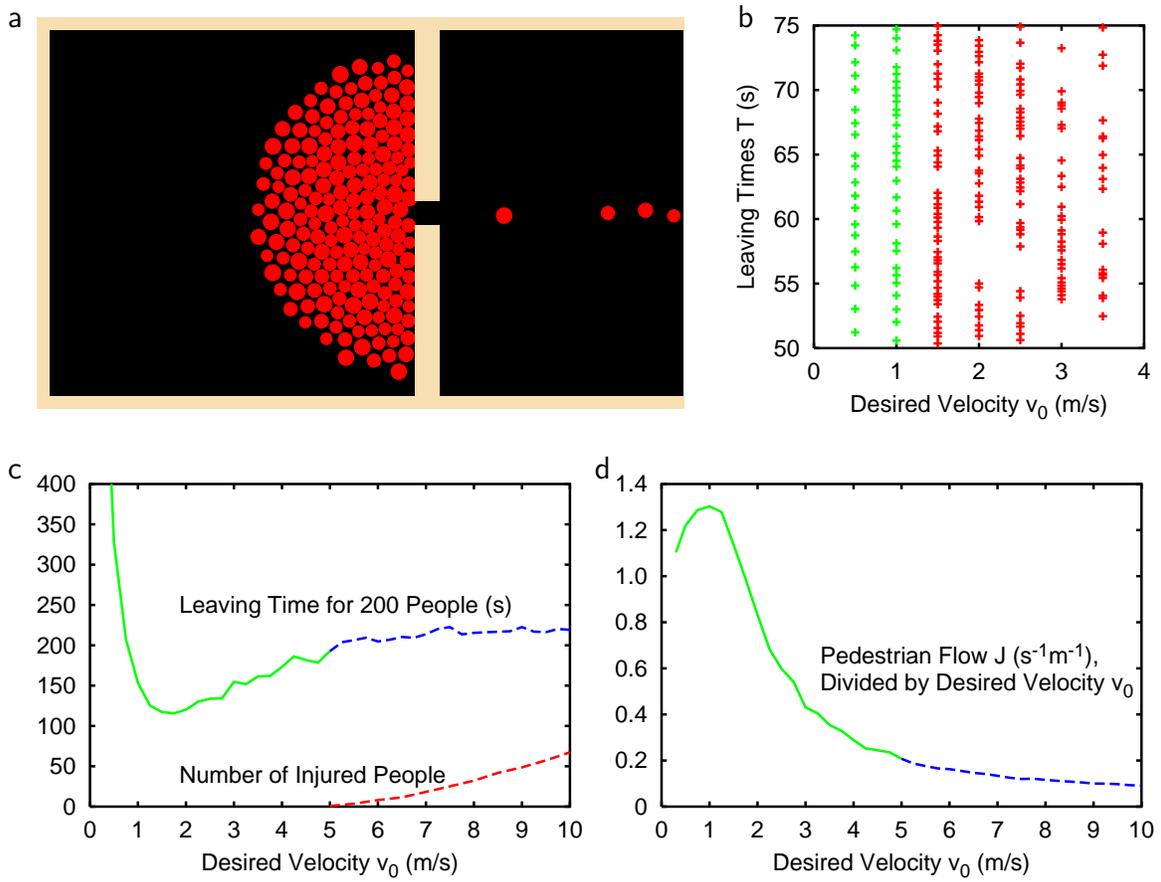
\unitlength1cm
\begin{figure}
\begin{center}
\begin{picture}(16,14)
\put(0,6){\includegraphics[height=6cm]{fig1a.cps}}
\put(9.5,11.9){\includegraphics[width=5.6cm, angle=-90]{fig1b.cps}}
\put(0,5.8){\includegraphics[width=5.6cm, angle=-90]{fig1c.cps}}
\put(7.6,5.8){\includegraphics[width=5.6cm, angle=-90]{fig1d.cps}}
\put(0,11.5){\sf a}
\put(9.7,11.5){\sf b}
\put(0,5.5){\sf c}
\put(7.8,5.5){\sf d}
\end{picture}
\end{center}
\caption[]{Simulation of pedestrians moving with identical
desired velocity $v_i^0=v_0$ towards the 1~m wide exit 
of a room of size 15~m\,$\times$\,15~m. 
{\sf a}~Snapshot of the scenario. Dynamic simulations are available at
{\tt http://angel.elte.hu/$\widetilde{\hphantom{n}}$panic/}.
{\sf b}~Illustration of leaving times of pedestrians for
various desired velocities $v_0$. Irregular outflow due to clogging is
observed for high desired velocities ($v_0 \ge 1.5$~m/s, see red plusses).
{\sf c}~Under conditions of normal walking, the time for 200
pedestrians to leave the room decreases with growing $v_0$.
Desired velocities higher than 1.5~m/s reduce the efficiency 
of leaving, which becomes particularly clear,
when the outflow $J$ is divided by the desired
velocity (see {\sf d}). This is due to
pushing, which causes additional friction effects. Moreover, above a
desired velocity of about $v_0 = 5$~m/s (--~--), people are
injured and become non-moving obstacles for others, 
if the sum of the magnitudes of the
radial forces acting on them divided by their circumference exceeds
a pressure of 1600~N/m.\cite{calibrate}
\\
Due to the above ``faster-is-slower effect'',
panics can be triggered by pedestrian counterflows,\protect\cite{Fire2}
which cause delays to the crowd
intending to leave. This makes the stopped pedestrians 
impatient and pushy. One may describe this by increasing
the desired velocity according to 
$v_i^0(t) = [1-p_i(t)]v_i^0(0) + p_i(t) v_i^{\rm max}$,
where $v_i^0(0)$ is the initial and $v_i^{\rm max}$
the maximum desired velocity. The time-dependent parameter
$p_i(t) = 1 - \overline{v}_i(t)/v_i^{\rm 0}$, where $\overline{v}_i(t)$
denotes the average speed into the desired direction of motion, is a measure of
impatience. 
Altogether, long waiting times increase the desired velocity,
which can produce inefficient outflow. This further increases
the waiting times, and so on, so that this tragic feedback can
eventually trigger panics. It is, therefore, imperative, to have sufficiently
wide exits and to prevent counterflows, when big crowds want to leave.
\label{fig2}}
\end{figure}

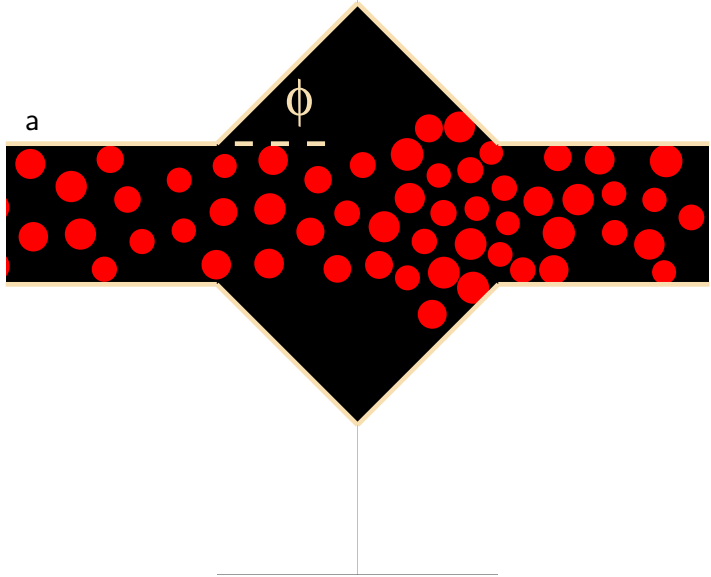
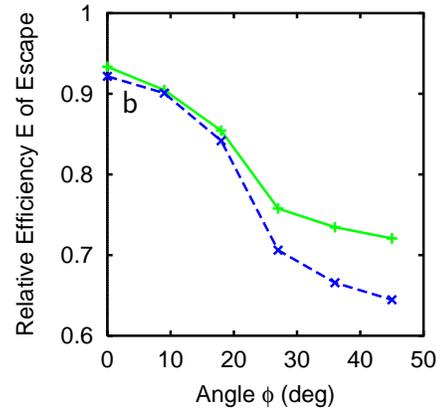
\begin{figure}
\begin{center}
\begin{picture}(16,7)
\put(-1,0.5){\includegraphics[height=6cm]{fig2a.cps}}
\put(9,6.2){\includegraphics[width=5.6cm, angle=-90]{fig2b.cps}}
\put(-0.5,4.5){\sf a}
\put(10.6,4.5){\sf b}
\end{picture}
\end{center}
\caption[]{Simulation of an escape route with a wider area (see also the Java
applets supplied at {\tt http://angel.elte.hu/$\widetilde{\hphantom{n}}$panic/}).
{\sf a} Illustration of the scenario with
$v_i^0 = v_0 = 2$~m/s. The corridor is 3~m wide and 15~m
long, the length of the triangular pieces in the middle
being 2$\times$3~m\,=\,6~m.
Pedestrians enter the simulation area on the left-hand side
with an inflow of $J=5.5$~s$^{-1}$m$^{-1}$
and flee towards the right-hand side.
{\sf b} Efficiency of leaving as a function of the angle $\phi$
characterising the width of the central zone, i.e., the
difference from a linear corridor. The relative efficiency
$E = \langle \vec{v}_i\cdot\vec{e}_i^0 \rangle /v_0$ measures the
average velocity along the corridor compared to the desired velocity
and lies between 0 and 1 (---).
While it is almost one (i.e., maximal) for a linear corridor ($\phi = 0$),
the efficiency drops by about 20\%, if the corridor contains a
widening. This becomes comprehensible, if we
take into account that the widening leads to disturbances by
pedestrians, who expand in the wide area due to their repulsive
interactions or try to overtake each other,
and squeeze into the main stream again at the end of the widening.
Hence, the right half of the illustrated corridor acts like a
bottleneck and leads to jamming. The drop of efficiency $E$ is even more pronounced,
(i) in the area of the widening where pedestrian flow is most irregular (--~--), 
(ii) if the corridor is
narrow, (iii) if the pedestrians have
different or high desired velocities, and (iv) if the pedestrian
density in the corridor is higher. 
\label{fig3}}
\end{figure}

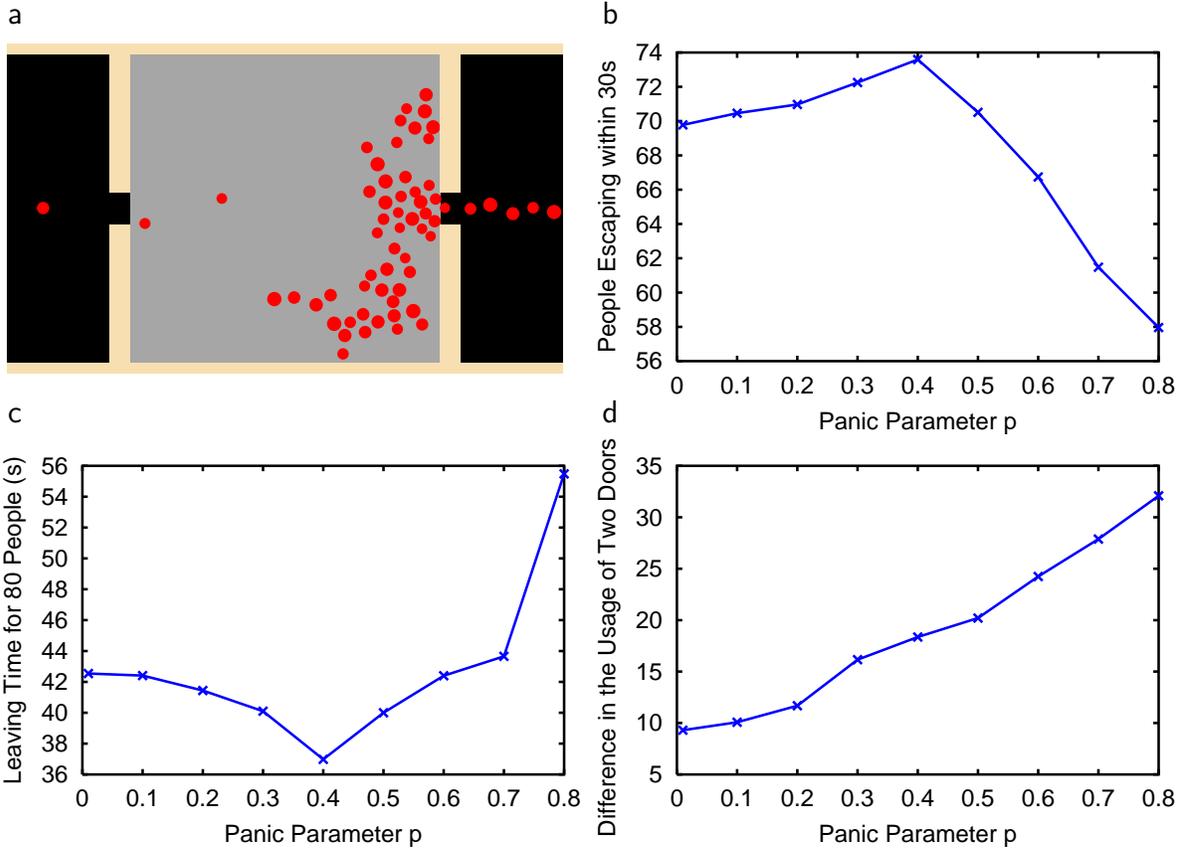
\begin{figure}
\begin{center}
\begin{picture}(16,11.5)
\put(-0.0,10.9){\includegraphics[height=4.35cm, angle=180]{fig3a.cps}}
\put(7.65,11.2){\includegraphics[width=5.45cm, angle=-90]{fig3b.cps}}
\put(-0.25,5.7){\includegraphics[width=5.45cm, angle=-90]{fig3c.cps}}
\put(7.65,5.7){\includegraphics[width=5.45cm, angle=-90]{fig3d.cps}}
\put(0,11.2){\sf a}
\put(7.9,11.2){\sf b}
\put(7.9,5.9){\sf d}
\put(0,5.9){\sf c}
\end{picture}
\end{center}
\caption[]{Simulation of $N=90$ pedestrians trying to escape a smoky room
of area $A=15\mbox{m}\,\times\,15\mbox{ m}$ (grey)
through two invisible doors of 1.5~m width,
which have to be found with a mixture
of individualistic and herding behaviour.  Java applets are available at
{\tt http://angel.elte.hu/$\widetilde{\hphantom{n}}$panic/}. {\sf a}~Snapshot of the
simulation with $v_i^0 = v_0 = 5$~m/s. 
Initially, each pedestrian selects his or her desired walking
direction randomly. Afterwards, a pedestrian's
walking direction is influenced by
the average direction of the neighbours
within a radius of, for example, $R_i=R=5$~m. 
The strength of this herding effect grows with increasing
panic parameter $p_i=p$ and increasing value of
$h=\pi R^2 \rho$, where $\rho = N/A$ denotes the pedestrian density.
When reaching a boundary, the direction of a pedestrian is
reflected. If one of the exits is closer than 2~m,
the room is left. {\sf b}~Number of people who manage to escape within 30~s
as a function of the panic parameter $p$.
{\sf c}~Illustration of the time required by 80 individuals to leave the
smoky room. If the exits are relatively narrow and
the degree $p$ of herding is small or large, leaving takes
particularly long, 
so that only some of the people escape before being poisoned by smoke.
Our results suggest that the best escape strategy is a certain compromise
between following of others and an individualistic searching
behaviour. This fits well into experimental data on the efficiency 
of group problem solving,\protect\cite{probsolv}
according to which groups normally perform better than
individuals, but masses are inefficient in finding new solutions to complex
problems. 
{\sf d}~Absolute difference $|N_1-N_2|$ in the numbers $N_1$ and $N_2$ of
persons leaving through the left exit or the right exit as a function of the
herding parameter $p$. We find that pedestrians tend to jam up at one
of the exits instead of equally using all available exits,
if the panic parameter is large.
\label{fig4}}
\end{figure}
\end{document}